# Mid-infrared optical properties of pyrolytic boron nitride in the 390–1050 °C temperature range using spectral emissivity measurements

I. González de Arrieta [a], T. Echániz [b,*], R. Fuente [c], L. del Campo [d], D. De Sousa Meneses [d], G.A. López [a], M.J. Tello [b,e]

[a] Dpto. Física Aplicada II, Facultad de Ciencia y Tecnología, Univ. del País Vasco UPV/EHU, Barrio Sarriena s/n, E-48940 Leioa, Bizkaia, Spain
[b] Dpto. Física de la Materia Condensada, Facultad de Ciencia y Tecnología, Univ. del País Vasco UPV/EHU, Barrio Sarriena s/n, E-48940 Leioa, Bizkaia, Spain
[c] Dpto. Matemática Aplicada, Escuela de Ingeniería de Bilbao, Univ. del País Vasco UPV/EHU, Alda. Urquijo s/n, E-48013 Bilbao, Spain
[d] CEMHTI-UPR3079, University of Orléans, CNRS Site Haute Température CS 90055, 1D Avenue de la Recherche Scientifique, 45071 Orléans Cedex 2, France
[e] Instituto de Síntesis y Estudio de Materiales, Univ. del País Vasco, Apdo. 644, E-48080 Bilbao, Spain

## ABSTRACT

This paper shows a systematic experimental and theoretical study on the temperature dependence of the infrared optical properties of pyrolytic boron nitride (pBN), from 390 to 1050 °C for wavelengths between 4 and 16 µm. The temperature dependence of these properties has never been analyzed before. The measured emissivity spectra were fitted to a dielectric function model and an effective medium ap-proximation. The phonon frequencies and dielectric constants agreed well with room temperature experimental values from the literature, as well as with *ab initio* and first principles calculations. In ad-dition, the phonon frequencies of the perpendicular mode and the dielectric constants of the parallel mode showed an appreciable parabolic temperature dependence, which justifies the interest of more theoretical efforts in order to explain this behavior. Finally, the results of this work demonstrate that thermal emission spectroscopy allows obtaining the values of the optical and dielectric parameters of impure ceramic materials in a simple manner as a function of temperature.

## 1. Introduction

Boron nitride shows four polymorphic variants whose special properties make this compound suitable for a large number of technological applications [1–4]. The hexagonal form (h-BN) is the most common phase and the one that shows the most unique electronic and optical properties. It is a very good electrical insulator, has great chemical stability and oxidation resistance, low thermal conductivity and can withstand extreme temperature conditions. Due to these properties, it is widely used in microelectronic devices, nuclear energy, X-ray lithography, lubrication, vacuum technology or high-performance oxidation-resistant coatings for metals and graphene up to 1100 °C in an oxidizing atmosphere [5–10]. Recently, the discovery of several new features has increased its interest. Firstly, the achievement of h-BN nanotubes with this simple hexagonal structure [11]. Secondly, its lasing capacity at a wavelength of 210 nm, which shows great potential for optoelectronic applications in the ultraviolet range [12]. Thirdly, its very good optical selectivity (infrared emitter and visible reflector) has made it a good candidate for coating heat shields of solar probes [13]. Finally, the presence of two phonon bands in the mid-infrared can be exploited to tune its optical properties by coupling them to plasmon polaritons from graphene or from a metal grating [14–16].

Hexagonal BN crystallizes in the $P6_3/mmc$ space group with four atoms in the unit cell and with the following irreducible representation of the phonon modes at the Brillouin zone center: $\Gamma_{vib} = A_{2u}(IR) + E_{1u}(IR) + 2E_{2g}(Raman) + 2B_{1g}(silent)$. This h-BN polymorphic modification consists of $sp^2$-bonded 2D layers [1]. Within each layer the atoms are bound by strong covalent bonds, whereas the layers are held together by weak Van der Waals forces. Since 1950, a continued effort has been made for the characterization of the physical properties of this compound. In particular, the optical properties of h-BN were studied by means of a large number of experimental techniques, mainly for wavenumbers above 3000 cm$^{-1}$, using single crystals, polycrystals and thin films deposited on different substrates (see for example references in [7]). First principles and *ab initio* calculations were made using different theories and approximations in order to analyze the experimental results. However, infrared and Raman spectroscopic experimental data that characterize the optical properties in the mid-infrared range or that allow obtaining the energies of the phonons at the $\Gamma$-point of the Brillouin zone are scarce [17–24]. The same scarcity occurs with respect to the number of publications that give theoretical values of the dielectric constants and the frequencies of the $\Gamma$-point phonons based on *ab initio* calculations [25–30]. Moreover, some discrepancies may be found among data from the literature, mainly in the parallel components of the static and high frequency dielectric constants, but also in some of the IR active phonon frequencies [17,19–21,24]. These discrepancies are also observed between experimental and theoretical values [25–30].

In this paper, we report an experimental study on the emissivity spectra of pyrolytic boron nitride (pBN), as well as its optical constants, $\Gamma$ phonon frequencies and dielectric properties (high-frequency and static dielectric constants) obtained by thermal emission spectroscopy. The theoretical interpretation of the experimental results is carried out by using a four-parameter dielectric function model as well as a Kramers–Kronig analysis. In addition to the basic interest of these measurements for the analysis of the discrepancies indicated above, this material shows promising properties in the mid-infrared range that make it attractive for many applications. As noted above, pBN may be an appropriate candidate for the main thermal protection system of the Solar Probe Plus (NASA) [10,31] or, from a broader viewpoint, as a high temperature passive thermal controller. Besides, tunable infrared absorption has also been achieved with a h-BN metamaterial, and it is known that numerical simulations for these applications require accurate values of the temperature-dependent dielectric function [32]. In order to implement these possible applications, it is necessary to expand the scarce number of published data on the temperature dependence of its optical properties [31,33,34]. The paper also presents the first study on the temperature dependence of the optical constants, $\Gamma$ phonon frequencies and dielectric properties (high-frequency and static dielectric constants) of pBN samples.

## 2. Experimental

The pBN sample studied in this paper consisted of a 1 mm thick square plate of hexagonal BN deposited on a graphite substrate by pyrolysis of precursor gases at 1800 °C [35]. The route employed ensures good homogeneity and high purity, since it does not involve binding phases. The composition and microstructure of the sample were characterized by chemical analysis, X-ray diffraction (XRD) and scanning electron microscopy (SEM). The X-ray diffractogram (shown in Fig. 1) and the SEM images are in complete agreement with those of a previous sample [24]. The sample density was measured to be 2.03 g/cm$^3$, a value 11% lower than the single-crystal value. The surface roughness parameters, measured with a conventional profilometer, are $R_a = 0.90$ µm, $R_q = 1.08$ µm, $R_z = 4.35$ µm and $R_t = 5.15$ µm.

The emissivity measurements were made with a radiometer which has been described in the literature [36]. The sample chamber was purged with dry air in order to avoid $CO_2$ and water absorption. The sample was heated with a $CO_2$ laser ($\lambda = 10.6$ µm), which allows reaching temperatures beyond 1000 °C. The sample thermal emission was collected with a Fourier transform infrared spectrometer calibrated by a black-body. The emissivity was then determined in the whole spectrum by comparison of the sample and black-body signals through the following formula:

$$E = \frac{FT(I_S - I_{RT})}{FT(I_{BB} - I_{RT})} \times \frac{P_{BB} - P_{RT}}{P_S - P_{RT}} E_{BB} \qquad (1)$$

where $FT$ means Fourier transform; $I_S$, $I_{BB}$ and $I_{RT}$ are the interferograms recorded for the sample, black-body and room temperature background of the apparatus; $P_S$, $P_{BB}$ and $P_{RT}$ are the Planck functions calculated at the temperatures of the sample, black-body and apparatus; and $E_{BB}$ is the emissivity of the black-body.

The temperature of the sample was determined at the Christiansen point ($\lambda_{Chris}$), a wavelength at which ceramic materials behave like a black-body ($E=1$). This point does not depend on temperature and is thus used as a common non-contact method for surface temperature measurement in ceramic materials [37]. In this paper, it was determined by a room-temperature diffuse reflectance measurement using an integrating sphere, with a value of $\lambda_{Chris} = 6.0$ µm.

## 3. Results and discussion

The X-ray diffractogram of Fig. 1 confirms the polycrystalline nature of the material, with no preferential orientation of the crystallites. It also reveals, together with the results from SEM microscopy, the presence of a small amount of graphite, which originates in the deposition process and has been found in similar samples in the literature [38]. The size of the graphite inclusions was observed to be smaller than the wavelength range of interest in this paper, which justifies the choice of the theoretical models for the interpretation of the data.

The emissivity spectrum of pBN was measured in the 390–1050 °C temperature range and for wavelengths between 4 and 16 µm. The wavelength range of study was limited by the onset of transmittance at both extremes, which makes the emissivity of the material to fall abruptly to 0. The maximum temperature was chosen to avoid any possible oxidation, while the minimum temperature chosen was the lowest one that ensured an acceptable signal-to-noise ratio in the selected spectral range. The emissivity spectra for three temperatures are shown in Fig. 2.

The spectra show two *reststrahlen* bands at 6–8 and 12–14 µm, corresponding to the two infrared active phonons of h-BN. These spectra show a great similarity to those observed by reflectance measurements at room temperature [17,24] and with the only emissivity spectrum found in the literature [34]. No experimental data in the infrared was found for single crystals. However, there are slight differences between the measured spectrum on polycrystalline samples [17] or highly oriented ones (hopBN) [24] with the measured spectrum in this paper. Group theory predicts that the observation of each phonon requires a determined direction of polarization ($\vec{E} \| c$ and $\vec{E} \perp c$). This occurs for hopBN samples but not for pBN, for which, as shown in Fig. 2, both phonons are observed in a near-normal measurement without polarization. This behavior is associated, as mentioned before, with the presence of randomly oriented crystallites observed from X-ray diffraction and SEM microscopy.

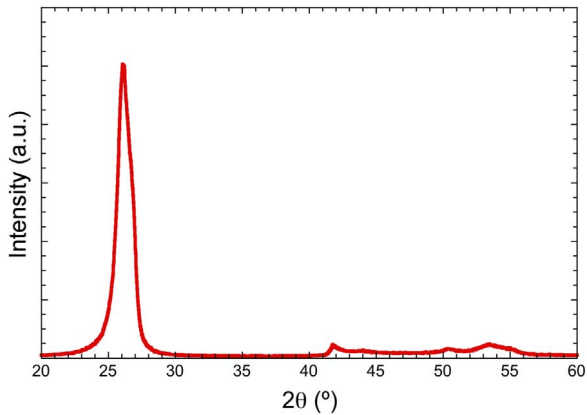

**Fig. 1.** X-ray diffractogram of the sample, taken with Cu-$K_\alpha$ radiation.

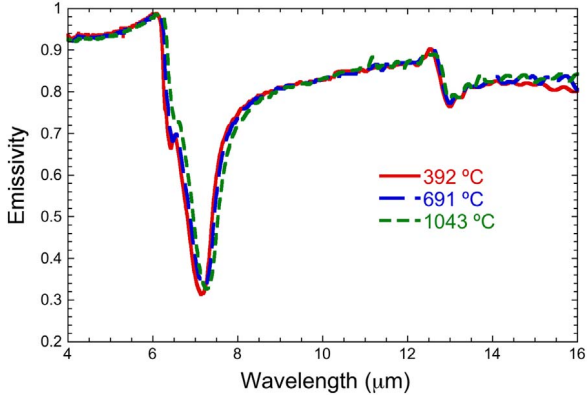

Fig. 2. Emissivity spectrum of pBN for three selected temperatures.

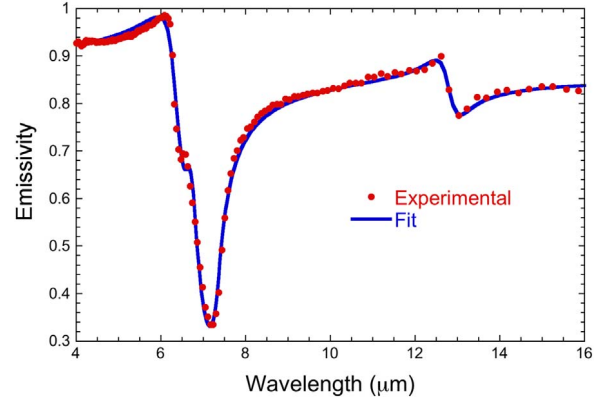

Fig. 3. Theoretical fit of the experimental data at 691 °C to Eqs. (2)–(4). Only some experimental points are shown.

A minor peak occurs at 6.4 μm, which becomes progressively weaker with temperature. This peak has been observed in reflectivity measurements on polycrystalline samples at room temperature and was attributed to 2-phonon combination processes [17]. However, more realistic recent results attribute this feature to the presence of pores and impurities in similar ceramic materials, such as NiO [39], SiC [40] and GaN [41].

The observed Christiansen wavelength in this sample was $\lambda_{Chris} = 6.0$ μm, a value that agrees with that obtained with a hopBN sample ($\lambda_{Chris} = 5.9$ μm) [24]. The characterization of this point is very useful for pyrometric temperature measurements in vacuum or in a dry atmosphere (as in this paper), while the measured value of $\varepsilon \simeq 0.93$ at the atmospheric window of 4 μm is more relevant for industrial applications of this material.

In order to analyze the pBN spectra, the data was fitted to a dielectric function model, which is related to the emissivity and reflectance through the Fresnel relations [42]. The model chosen, called the four-parameter or Kurosawa model [43], has been successfully employed to fit optical spectra of a great number of ceramic materials [44,45]:

$$\epsilon_{\perp,\parallel}(\omega) = \epsilon_\infty \frac{\Omega_{LO\perp,\parallel}^2 - \omega^2 + i\gamma_{LO\perp,\parallel}\omega}{\Omega_{TO\perp,\parallel}^2 - \omega^2 + i\gamma_{TO\perp,\parallel}\omega} \quad (2)$$

where $\epsilon_\infty$ is the high-frequency electronic contribution to the dielectric function, $\Omega$ is the frequency of an optical phonon (transverse or longitudinal), $\gamma$ is the damping frequency of that phonon, and $\perp$ and $\parallel$ are the two possible electric field polarizations for a hexagonal crystal (perpendicular or parallel to the optical c-axis).

As the sample is a randomly oriented polycrystal, the global dielectric function should be an average of the two polarizations, in which the perpendicular accounts for 2/3 of the total because of its higher multiplicity:

$$\epsilon_{av} = \frac{1}{3}\epsilon_\parallel + \frac{2}{3}\epsilon_\perp \quad (3)$$

Finally, the effect of the microstructure and the presence of carbon impurities observed by the XRD and SEM experimental techniques, in agreement with data in the literature for similar samples, should be taken into account. To this end, and given the small size of the voids and defects ($< 1$ μm), the Maxwell–Garnett effective medium theory was applied [46]. This model takes into account the presence of spherical pores of dielectric function $\epsilon_i$ and its volume fraction $f_i$, and returns an effective value of the dielectric function of the global medium:

$$\frac{\epsilon_{ef} - \epsilon_{av}}{\epsilon_{ef} + 2\epsilon_{av}} = f_i \left( \frac{\epsilon_i - \epsilon_{av}}{\epsilon_i + 2\epsilon_{av}} \right) \quad (4)$$

where $\epsilon_{ef}$ stands for the effective dielectric function, $\epsilon_i$ for one of the impurities, $f_i$ for their volume concentration and $\epsilon_{av}$ for the average dielectric function of the polycrystal, given by Eq. (3).

The experimental data for an intermediate temperature of 691 °C have been fitted to this model, with the result shown in Fig. 3. There is an excellent agreement, even around the 6.4 μm minor peak, which could not be accounted for without the effective medium theory.

The parameters obtained with this model (phonon frequencies and dielectric constants) are shown in Table 1, together with other literature experimental and theoretical parameters obtained using first principles calculations. The temperature dependence of these parameters is plotted in Figs. 4 and 5. A first view of Table 1 indicates that there is a small variation between the values of thin films and those of bulk samples. Among the bulk samples, the only noticeable discrepancy appears in the values of the dielectric constants obtained with a polycrystalline sample [17]. These data were the only experimental values available until 1999 and disagree with all theoretical predictions prior to this date [25,26,28]. Due to the need of an accurate prediction of the band-gap for first principles calculations, it was argued that the discrepancy could be associated with the difference between the calculated and measured values for the gap. For the rest of the experimental values of the parameters of Table 1, the discrepancies (<1.5%) are within the experimental error, even for thin films. Besides, the values of the dielectric constants are insensitive to the method used (e.g., ultra soft or norm-conserving pseudopotentials). However, the same is not true for all phonon frequencies. All theoretical predictions give values of $\Omega_{TO,\parallel}$ between 3% and 5% lower than those found experimentally. For the other modes there is a slight discrepancy between the theoretical values, which are reasonably close to the experimental ones. The only exception is the $\Omega_{LO,\perp}$ in Reference [27].

**Table 1**
Compilation of experimental and theoretical values of the phonon frequencies (in cm$^{-1}$) and dielectric constants of pBN. The values shown for this paper correspond to the lowest temperature measured (392 °C) to allow comparison with room temperature values. All theoretical calculations referenced rely on the local density approximation (LDA) of the density functional theory. The kind of atomic bases or pseudopotentials used are specified in the first column.

| Sample | $\Omega_{TO,\parallel}$ | $\Omega_{LO,\parallel}$ | $\Omega_{TO,\perp}$ | $\Omega_{LO,\perp}$ | $\epsilon_{0,\perp}$ | $\epsilon_{0,\parallel}$ | $\epsilon_{\infty,\perp}$ | $\epsilon_{\infty,\parallel}$ | Ref. |
|---|---|---|---|---|---|---|---|---|---|
| *Experimental* | | | | | | | | | |
| This paper | 773 | 829 | 1373 | 1601 | 7.06 | 2.52 | 5.19 | 2.19 | |
| hopBN | 770 | 826 | 1365 | 1622 | 6.3 | 2.3 | 4.4 | 2.0 | [24] |
| pBN | 783 | 828 | 1367 | 1610 | 7.04 | 5.09 | 4.95 | 4.1 | [17] |
| pBN | 770 | – | 1383 | – | – | – | – | – | [20] |
| Thin film (sputtering) | 766 | 811 | 1400 | 1586 | – | – | – | – | [21] |
| Thin film (PECVD) | 790 | 828 | 1386 | 1599 | – | – | 4.71 | 2.72 | [22] |
| *Theory* | | | | | | | | | |
| OLCAO | – | – | – | – | – | – | 4.32 | 2.21 | [25] |
| Ultrasoft | 754 | 823 | 1382 | 1614 | 6.61 | 3.38 | 4.85 | 2.84 | [26] |
| Ultrasoft | 757 | 783 | 1366 | 1587 | – | – | – | – | [27] |
| Norm-conserving | 746 | 819 | 1372 | 1610 | 6.71 | 3.57 | 4.87 | 2.95 | [28] |
| Norm-conserving | 745 | – | 1375 | 1607 | – | – | – | – | [29] |
| Norm-conserving | 760 | 824 | 1376 | 1610 | – | – | – | – | [30] |

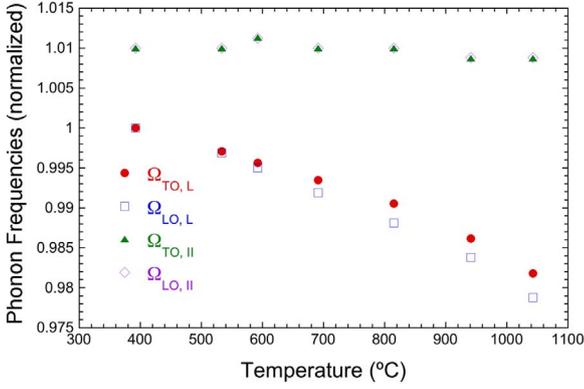

**Fig. 4.** Temperature dependence of the phonon frequencies, normalized to the value for the lowest temperature (indicated in Table 1). The parallel components have been shifted 0.01 upwards for clarity.

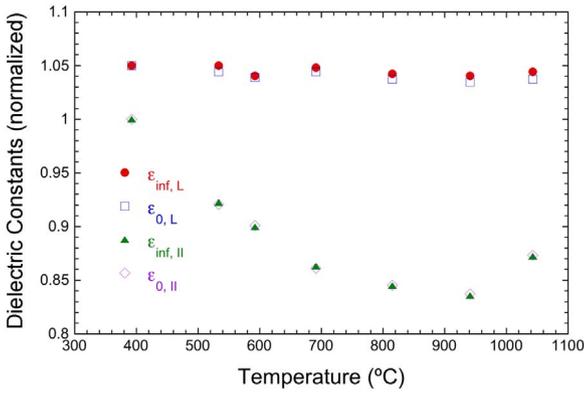

**Fig. 5.** Temperature dependence of the static and high-frequency dielectric constants, normalized to the value for the lowest temperature (indicated in Table 1). The perpendicular components have been shifted 0.05 upwards for clarity.

The other parameters of the Kurosawa model have not been thoroughly studied. The errors introduced by the fitting procedure in the values of the damping constants are greater than the slight positive temperature dependence that could be observed. The mean values for the perpendicular modes are $\gamma_{TO,\perp} = 42 \pm 2$ cm$^{-1}$ and $\gamma_{LO,\perp} = 99 \pm 8$ cm$^{-1}$. No conclusions could be extracted from the parallel damping constants, since the fitting algorithm introduced noticeable errors near the end of the spectral range. The values reported above are in good agreement with the data given in the literature for room temperature [22,24]. The effective medium parameters have been found to be equal at all temperatures, within the expected uncertainty and with values of $\epsilon_i = 5.0$ and $f_i = 8\%$. These parameters are crucial for an accurate description of the global emission spectra, but the approximation chosen is not precise enough to allow any discussion of the properties of the impurities themselves.

The temperature dependence of the phonon frequencies and dielectric constants are plotted in Figs. 4 and 5. All data in the figures are normalized to the value at 392 °C (given in Table 1) and shifted for clarity. Two facts should be highlighted. The frequencies of the perpendicular mode and the dielectric constants of the parallel one show an appreciable temperature dependence. On the contrary, the other components show a very slight dependence. The parallel dielectric constants show a parabolic behavior that passes through a minimum around 900 °C, whereas for the perpendicular components of the frequencies the behavior is concave downwards. No explanation was found for the behavior of the dielectric constant, but the perpendicular frequencies show the typical shape due to three- and four-phonon interactions [47]. This highlights the need for theoretical calculations that take temperature into account, as well as more experimental studies of the temperature-dependent optical parameters in other ceramic compounds.

In order to obtain more precise values of the dielectric function of the material, a Kramers–Kronig inversion of the reflectivity of the sample ($R = 1 - E$) was performed. Contrary to the model-fitting approach, this method involves purely mathematical operations on the data without any physical input. The imaginary part of the Fresnel reflection coefficient ($r^* = re^{i\theta}$) is calculated from the real part ($r = \sqrt{R}$) by means of the following equation [48]:

$$\theta(\omega) = -\frac{1}{\pi} P \int_0^\infty \ln\left|\frac{s+\omega}{s-\omega}\right| \frac{d(\ln r(s))}{ds} ds \qquad (5)$$

where $P$ stands for principal value of the integral.

The result of the Kramers–Kronig is shown in Fig. 6, together with the dielectric functions generated with the parameters contained in Table 1.

It can be seen that the two methods lead to curves which overlap for most of the spectral range considered. However, there is a noticeable difference around the main phonon at 7.5 µm, more clearly seen in the imaginary part of the dielectric function. This may reveal an inadequacy of the semiclassical models when trying to obtain precise values of the optical constants of materials, even when they reproduce successfully the experimental emissivity or reflectivity data. A similar problem was observed in a pBN sample from the literature, for which only the effect of porosity ($\epsilon_i = 1$) was considered and not that of general impurities [24]. The modelling present on this paper improves the quality of the fitting, and thus the discrepancies between the dielectric functions are lower than in that previous paper, but still significant. However, no definitive conclusions can be made in the absence of an independent determination of the optical constants.

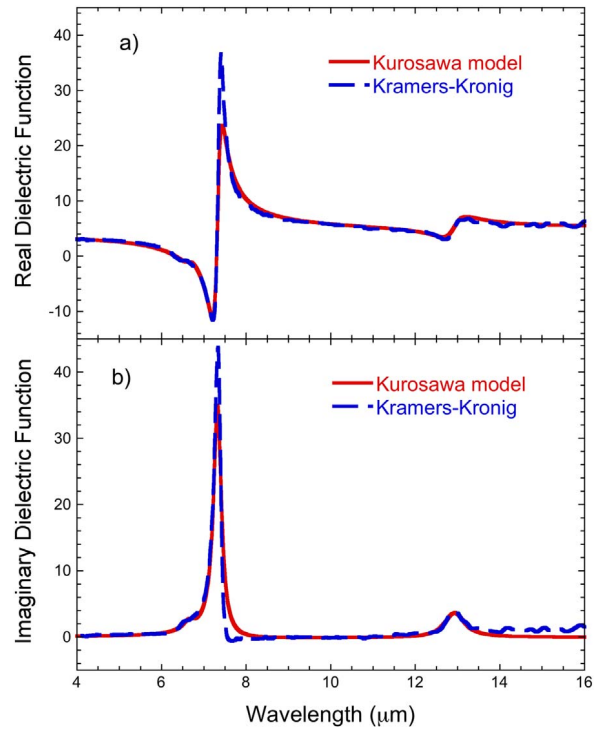

**Fig. 6.** Comparison of the real (a) and imaginary (b) parts of the dielectric functions obtained from the experimental emissivity data at 691 °C from the fitting of the Kurosawa model and the Kramers–Kronig inversion.

## 4. Conclusions

The emissivity spectra of pBN were measured as a function of temperature and fitted to a four-parameter dielectric function model taking into account the random orientation of the crystals and the presence of carbon impurities, which were described by an effective medium approximation. This modelling approach shows a better agreement between the theoretical and experimental spectra than previous attempts.

The temperature dependence of the optical and dielectric parameters of this material has been reported for the first time. The values obtained by the fitting procedure agree well with experimental data at room temperature and with first-principles calculations. No systematic differences have been found for different sample types, except for those of thin films. A parabolic temperature dependence was found for the frequencies of the perpendicular components of the phonons and for the parallel components of the dielectric constants. This temperature dependence could not be explained by the current *ab initio* calculations, which rely almost exclusively on the $T=0$ approximation.

The quality of the experimental results presented in this work confirms that thermal emission spectroscopy is a good alternative for the study of optical properties of ceramics beyond room temperature. Knowledge of the emissivity spectra at high temperatures is necessary for many interesting applications, such as protective and selective coatings for the aerospace and solar energy industries. Moreover, accurate temperature-dependent values of the dielectric function are also crucial for numerical simulations of complex materials based on h-BN, such as tunable infrared absorbers.

## Acknowledgments

This research was partially supported by the research program of UPV/EHU. T. Echániz would like to acknowledge the UPV/EHU for its support through a postdoctoral fellowship. I. González de Arrieta thanks the Radiative and Transport Properties Group at CEMHTI laboratory (CNRS) in Orléans for the opportunity to use its facilities and the useful discussions.

## References

[1] Edgar J. Properties of group III nitrides. London: Inspect; 1994.
[2] Duan X, Yang Z, Chen L, Tian Z, Cai D, Wang Y, et al. Review on the properties of hexagonal boron nitride matrix composite ceramics. J Eur Ceram Soc 2016;36:3725–3737. http://dx.doi.org/10.1016/j.jeurceramsoc.2016.05.007.
[3] Engler M, Lesniak C, Damasch R, Ruising B, Eichler J. Hexagonal boron nitride (hBN) – applications from metallurgy to cosmetics. CFI-Ceram Forum Int 2007;84:49–53.
[4] Dean CR, Young AF, Meric I, Lee C, Wang L, Sorgenfrei S, et al. Boron nitride substrates for high-quality graphene electronics. Nat Nanotechnol 2010;5:722–726. http://dx.doi.org/10.1038/nnano.2010.172.
[5] Solozhenko V, Lazarenko A, Petitet J-P, Kanaev A. Bandgap energy of graphite-like hexagonal boron nitride. J Phys Chem Solids 2001;62:1331–1334. http://dx.doi.org/10.1016/S0022-3697(01)00030-0.
[6] Widmayer P, Boyen H-G, Ziemann P, Reinke P, Oelhafen P. Electron spectroscopy on boron nitride thin films: comparison of near-surface to bulk electronic properties. Phys Rev B 1999;59:5233–5241. http://dx.doi.org/10.1103/PhysRevB.59.5233.
[7] Liu L, Feng YP, Shen ZX. Structural and electronic properties of h-BN. Phys Rev B 2003;68:104102. http://dx.doi.org/10.1103/PhysRevB.68.104102.
[8] Pierson HO. Handbook of chemical vapor deposition (CVD).Norwich, NY: William Andrew Publishing; http://dx.doi.org/10.1016/B978-081551432-9.50001-2.
[9] Liu Z, Gong Y, Zhou W, Ma L, Yu J, Idrobo J, et al. Ultrathin high-temperature oxidation-resistant coatings of hexagonal boron nitride. Nat Commun 2013;4:2541. http://dx.doi.org/10.1038/ncomms3541.
[10] Balat-Pichelin M, Eck J, Heurtault S, Glénat H. Experimental study of pyrolytic boron nitride at high temperature with and without proton and VUV irradiations. Appl Surf Sci 2014;314:415–425. http://dx.doi.org/10.1016/j.apsusc.2014.07.007.
[11] Chopra NG, Luyken RJ, Cherrey K, Crespi VH, Cohen ML, Louie SG, et al. Boron nitride nanotubes. Science 1995;269:966–967. http://dx.doi.org/10.1126/science.269.5226.966.
[12] Watanabe K, Taniguchi T, Kanda H. Direct-bandgap properties and evidence for ultraviolet lasing of hexagonal boron nitride single crystal. Nat Mater 2004;3:404–409. http://dx.doi.org/10.1038/nmat1134.
[13] Brodu E, Balat-Pichelin M, de Sousa Meneses D, Sans J-L. Reducing the temperature of a C/C composite heat shield for solar probe missions with an optically selective semi-transparent pyrolytic boron nitride (pBN) coating. Carbon 2015;82:39–50. http://dx.doi.org/10.1016/j.carbon.2014.10.022.
[14] Zhao B, Zhang ZM. Perfect mid-infrared absorption by hybrid phonon-plasmon polaritons in hBN/metal-grating anisotropic structures. Int J Heat Mass Transf 2017;106:1025–1034. http://dx.doi.org/10.1016/j.ijheatmasstransfer.2016.10.074.
[15] Zhao B, Zhang Z. Enhanc photon tunneling surf plasmon–phonon polaritons graphene/hBN heterostruct. J Heat Transf 2017;139:022701. http://dx.doi.org/10.1115/1.4034793.
[16] Dai S, Ma Q, Liu M, Andersen T, Fei Z, Goldflam M, et al. Graphene on hexagonal boron nitride as a tunable hyperbolic metamaterial. Nat Nanotechnol 2015;10:682–686. http://dx.doi.org/10.1038/nnano.2015.131.
[17] Geick R, Perry CH, Rupprecht G. Normal modes in hexagonal boron nitride. Phys Rev 1966;146:543–547. http://dx.doi.org/10.1103/PhysRev.146.543.
[18] Kuzuba T, Era K, Ishii T, Sato T. A low frequency Raman-active vibration of hexagonal boron nitride. Solid State Commun 1978;25:863–865. http://dx.doi.org/10.1016/0038-1098(78)90288-0.
[19] Nemanich R, Solin S, Martin RM. Light scattering study of boron nitride microcrystals. Phys Rev B 1981;23:6348. http://dx.doi.org/10.1103/PhysRevB.23.6348.
[20] Hoffman DM, Doll GL, Eklund PC. Optical properties of pyrolytic boron nitride in the energy range 0.05–10 eV. Phys Rev B 1984;30:6051–6056. http://dx.doi.org/10.1103/PhysRevB.30.6051.
[21] Schubert M, Rheinländer B, Franke E, Neumann H, Tiwald TE, Woollam JA, et al. Infrared optical properties of mixed-phase thin films studied by spectroscopic ellipsometry using boron nitride as an example. Phys Rev B 1997;56:13306. http://dx.doi.org/10.1103/physrevb.56.13306.
[22] Ben el Mekki M, Mestres N, Pascual J, Polo M, Andújar J. Infrared and Raman analysis of plasma CVD boron nitride thin films. Diam Relat Mater 1999;8:398–401. http://dx.doi.org/10.1016/S0925-9635(98)00413-0.
[23] Reich S, Ferrari AC, Arenal R, Loiseau A, Bello I, Robertson J. Resonant Raman scattering in cubic and hexagonal boron nitride. Phys Rev B 2005;71:205201. http://dx.doi.org/10.1103/PhysRevB.71.205201.
[24] De Sousa Meneses D, Balat-Pichelin M, Rozenbaum O, del Campo L, Echegut P. Optical indices and transport scattering coefficient of pyrolytic boron nitride: a natural thermal barrier coating for solar shields. J Mater Sci 2016;51:4660–4669. http://dx.doi.org/10.1007/s10853-016-9781-2.
[25] Xu Y-N, Ching WY. Calculation of ground-state and optical properties of boron nitrides in the hexagonal, cubic, and wurtzite structures. Phys Rev B 1991;44:7787–7798. http://dx.doi.org/10.1103/PhysRevB.44.7787.
[26] Ohba N, Miwa K, Nagasako N, Fukumoto A. First-principles study on structural, dielectric, and dynamical properties for three BN polytypes. Phys Rev B 2001;63:115207. http://dx.doi.org/10.1103/PhysRevB.63.115207.
[27] Yu WJ, Lau WM, Chan SP, Liu ZF, Zheng QQ. *Ab initio* study of phase transformations in boron nitride. Phys Rev B 2003;67:014108. http://dx.doi.org/10.1103/PhysRevB.67.014108.
[28] Cai Y, Zhang L, Zeng Q, Cheng L, Xu Y. Infrared reflectance spectrum of BN calculated from first principles. Solid State Commun 2011;141:262–266. http://dx.doi.org/10.1016/j.ssc.2006.10.040.
[29] Serrano J, Bosak A, Arenal R, Krisch M, Watanabe K, Taniguchi T, et al. Vibrational properties of hexagonal boron nitride: inelastic x-ray scattering and *ab initio* calculations. Phys Rev Lett 2007;98:095503. http://dx.doi.org/10.1103/PhysRevLett.98.095503.
[30] Hamdi I, Meskini N. Ab initio study of the structural, elastic, vibrational and thermodynamic properties of the hexagonal boron nitride: performance of LDA and GGA. Physica B 2010;405:2785–2794. http://dx.doi.org/10.1016/j.physb.2010.03.070.
[31] Brodu E, Balat-Pichelin M. Emissivity of boron nitride and metals for the Solar Probe Plus Mission. J Spacecr Rocket 2016:1–9. http://dx.doi.org/10.2514/1.A33453.
[32] Hervé A, Drévillon J, Ezzahri Y, Joulain K, De Sousa Meneses D, Hugonin J-P. Temperature dependence of a microstructured SiC coherent thermal source. J Quant Spectrosc Radiat Transf 2016;180:29–38. http://dx.doi.org/10.1016/j.jqsrt.2016.04.007.
[33] Manara J, Caps R, Ebert H, Hemberger F, Fricke J, Seidl A. Infrared optical properties of semitransparent pyrolytic boron nitride (pBN). High Temp-High Press 2002;34:65–72. http://dx.doi.org/10.1068/htwu309.

[34] Manara J, Keller M, Kraus D, Arduini-Schuster M. Determining the transmittance and emittance of transparent and semitransparent materials at elevated temperatures. In: Proceedings of the 5th European Thermal-Sciences Conference, Eindhoven, 2008.
[35] Matsuda T, Uno N, Nakae H. Synthesis and structure of chemically vapour-deposited boron nitride. J Mater Sci 1986;21:649–658. http://dx.doi.org/10.1007/BF01145537.
[36] De Sousa Meneses D, Melin P, del Campo L, Cosson L, Echegut P. Apparatus for measuring the emittance of materials from far infrared to visible wavelengths in extreme conditions of temperature. Infrared Phys Technol 2015;69:96–101. http://dx.doi.org/10.1016/j.infrared.2015.01.011.
[37] Rousseau B, Brun JF, De Sousa Meneses D, Echegut P. Temperature measurement: Christiansen wavelength and blackbody reference. Int J Thermophys 2005;26:1277–1286. http://dx.doi.org/10.1007/s10765-005-6726-4.
[38] Naftaly M, Leist J, Fletcher JR. Optical properties and structure of pyrolytic boron nitride for THz applications. Opt Mater Express 2013;3:260–269. http://dx.doi.org/10.1364/OME.3.000260.
[39] Nikolić M, Blagojević V, Paraskevopoulos K, Zorba T, Vasiljević-Radović D, Nikolić P, et al. Far infrared properties of sintered NiO. J Eur Ceram Soc 2007;27:469–474. http://dx.doi.org/10.1016/j.jeurceramsoc.2006.04.065.
[40] Okamoto Y, Ordin SV, Kawahara T, Fedorov MI, Miida Y, Miyakawa T. Infrared-reflection characterization of sintered SiC thermoelectric semiconductors with the use of a four-component effective medium model. J Appl Phys 1999;85:6728–6737. http://dx.doi.org/10.1063/1.370186.
[41] Zhang X, Hou Y-T, Feng Z-C, Chen J-L. Infrared reflectance of GaN films grown on Si(001) substrates. J Appl Phys 2001;89:6165–6170. http://dx.doi.org/10.1063/1.1368162.
[42] Siegel R, Howell J. Thermal radiation heat transfer. New York: Taylor and Francis; 2002.
[43] Kurosawa T. Polarization waves in solids. J Phys Soc Jpn 1961;16:1298–1308. http://dx.doi.org/10.1143/JPSJ.16.1298.
[44] Gervais F. Optical conductivity of oxides. Mater Sci Eng R 2002;39:29–92. http://dx.doi.org/10.1016/S0927-796X(02)00073-6.
[45] González de Arrieta I, Echániz T, Pérez-Sáez R, Tello M. Thermo-radiative and optical properties of a cutting tool based on polycrystalline cubic boron nitride (PCBN). Mater Res Express 2016;3:045904. http://dx.doi.org/10.1088/2053-1591/3/4/045904.
[46] Wang M, Pan N. Predictions of effective physical properties of complex multiphase materials. Mater Sci Eng R 2008;63:1–30. http://dx.doi.org/10.1016/j.mser.2008.07.001.
[47] Gasanly N, Özkan H, Aydinli A, Yilmaz I. Temperature dependence of the Raman-active phonon frequencies in indium sulfide. Solid State Commun 1999;110:231–236. http://dx.doi.org/10.1016/S0038-1098(99)00062-9.
[48] De Sousa Meneses D, Rousseau B, Echegut P, Simon P. Retrieval of linear optical functions from finite range spectra. Appl Spectrosc 2007;61:1390–1397. http://dx.doi.org/10.1366/000370207783292163.